\documentclass[compsoc,conference,a4paper,10pt,times]{IEEEtran}
\IEEEoverridecommandlockouts
\usepackage{cite}
\usepackage{amsmath,amssymb,amsfonts}
\usepackage{algorithmic}
\usepackage{graphicx}
\usepackage{textcomp}
\usepackage{bmpsize}
\usepackage{xcolor}
\usepackage{lipsum}
\usepackage[colorlinks=true,urlcolor=black]{hyperref}
\def\BibTeX{{\rm B\kern-.05em{\sc i\kern-.025em b}\kern-.08em
    T\kern-.1667em\lower.7ex\hbox{E}\kern-.125emX}}

\usepackage[font=footnotesize]{caption}
\usepackage{subcaption}
\usepackage{enumitem,amssymb}
\usepackage{xspace}
\usepackage[frozencache,cachedir=minted-cache]{minted}
\usepackage{import}
\usepackage{multirow}
\usepackage{macros}
\begin{document}

\title{A Case Study on the Use of Representativeness Bias as a Defense Against Adversarial Cyber Threats}

\author{
\IEEEauthorblockN{Briland Hitaj}
\IEEEauthorblockA{
\textit{SRI International}\\
briland.hitaj@sri.com}
\and
\IEEEauthorblockN{Grit Denker}
\IEEEauthorblockA{
\textit{SRI International}\\
grit.denker@sri.com}
\and
\IEEEauthorblockN{Laura Tinnel}
\IEEEauthorblockA{
\textit{SRI International}\\
laura.tinnel@sri.com}
\and
\IEEEauthorblockN{Michael McAnally}
\IEEEauthorblockA{\textit{Two Six Technologies}\\
michael.mcanally@twosixtech.com}
\and
\IEEEauthorblockN{Bruce DeBruhl}
\IEEEauthorblockA{
\textit{SRI International}\\
bruce.debruhl@sri.com}
\and
\IEEEauthorblockN{Nathan Bunting}
\IEEEauthorblockA{\textit{Two Six Technologies}\\
nathan.bunting@twosixtech.com}
\and
\IEEEauthorblockN{Alex Fafard}
\IEEEauthorblockA{\textit{Two Six Technologies}\\
alex.fafard@twosixtech.com}
\and
\IEEEauthorblockN{Daniel Aaron}
\IEEEauthorblockA{\textit{Two Six Technologies}\\
daniel.aaron@twosixtech.com}
\and
\IEEEauthorblockN{Richard D. Roberts}
\IEEEauthorblockA{\textit{RAD Science Solution}\\
richard@radssolution.com}
\and
\IEEEauthorblockN{Joshua Lawson}
\IEEEauthorblockA{
\textit{SRI International}\\
joshua.lawson@sri.com}
\and
\IEEEauthorblockN{Greg McCain}
\IEEEauthorblockA{
\textit{SRI International}\\
greg.mccain@sri.com}
\and
\IEEEauthorblockN{Dylan Starink}
\IEEEauthorblockA{
\textit{SRI International}\\
dylan.starink@sri.com}
\thanks{To appear in the Proceedings of the $4^{th}$ Workshop on Active Defense and Deception (AD\&D), co-located with the $10^{th}$ IEEE European Symposium on Security and Privacy (EuroS\&P 2025)}
}

\maketitle

\thispagestyle{plain}
\pagestyle{plain}

\begin{abstract}
Cyberspace is an ever-evolving battleground involving adversaries seeking to circumvent existing safeguards and defenders aiming to stay one step ahead by predicting and mitigating the next threat. Existing mitigation strategies have focused primarily on solutions that consider software or hardware aspects, often ignoring the human factor. This paper takes a first step towards psychology-informed, active defense strategies, where we target biases that human beings are susceptible to under conditions of uncertainty. 

Using capture-the-flag events, we create realistic 
challenges that tap into a particular cognitive bias: representativeness. This study finds that this bias can be triggered to thwart hacking attempts and divert hackers into non-vulnerable attack paths. Participants were exposed to two different challenges designed to exploit representativeness biases. One of the representativeness challenges significantly thwarted attackers away from vulnerable attack vectors and onto non-vulnerable paths, signifying an effective bias-based defense mechanism.
This work paves the way towards cyber defense strategies that leverage additional human biases to thwart future, sophisticated adversarial attacks. 
\end{abstract}

\section{Introduction}
\label{sec:introduction}
Traditionally, cyberattack responses have been heavily based on the presence and use of antivirus software, intrusion detection systems (IDS), endpoint detection and response (EDR), and similar technologies. Although undoubtedly beneficial, 
these mechanisms are designed to facilitate post-attack responses, drastically hindering the defenders' ability to react. This has led to a paradigm shift from \emph{reactive} defense strategies to \emph{proactive} ones. Moving target defense approaches (also known as active defense)~\cite{pagnotta2023dolos, de2016ahead, ZhuangEtAl14} in conjunction with deception techniques (e.g., honeypots, honeynets)~\cite{Spitzner03, kahlhofer2024application, sladic2024llm, srinivasa2022deceptive} have helped to shift the scales significantly in favor of the defenders. 
However, such mechanisms are often insufficient.
Once an adversary is aware of their presence, they can rapidly change their strategies to account for each defense technique. 

In recent years, a growing body of work has sought to complement active defense (and deception) strategies. Particularly prominent has been research that seeks to account for behavioral patterns in adversary decision-making, paving the way toward stronger, more effective, and alluring honeypots~\cite{pappa2024modeling, sharma2023well, quibell2024towards, johnson2022decision, ferguson2023cyber, johnson2024adversarial, barron2021click, barron2020addressing, gutzwiller2024exploratory, gutzwiller2019cyber, johnson2022investigating, johnson2020cyber, kahlhofer2024honeyquest}. 
Under this broad framework, this paper seeks to expand the defenders' arsenal by focusing on a specific domain of cognitive psychology: biases and heuristics. 
In theory, cognitive biases can lead attackers to make predictable errors, creating opportunities for tailored defense strategies. Our work investigates the impact and influence of cognitive biases in an adversary's decision-making process, ultimately driving the research and development of novel cyber defenses tailored to an adversary's inherent cognitive biases. We pick a small subset of cognitive biases as proof-of-concept, all linked to representativeness bias. More specifically, for this work, we ask the following research questions:
\begin{itemize}
    \item \emph{What is the impact (and influence) of \textbf{representativeness bias (and its underlying facets)} in the decision-making process of an adversary?} 
    \item \emph{What measurable effect on attackers' cyber behavior occurs when a defense mechanism targets attackers' propensity to representativeness bias?}
\end{itemize}

Representativeness bias is a mental ``shortcut'' (i.e., heuristic) that people use when estimating probabilities. It is the tendency to judge the probability or frequency of a hypothesis by considering how much the hypothesis resembles available data (such as the stereotype or prototype they already have in mind)~\cite{BarHillel80, GoodieFantino95, StolarzFantinoEtAl06, TverskyKahneman82, Heuer99, GertnerEtAl16}.
Representativeness bias can be seen as a high-level bias that is realized by various lower-order facets. A pertinent example of this bias can be seen in users who rely on visual representation of security cues, such as padlock icons or HTTPS URLs, to assess the trustworthiness of a website~\cite{fard2022role}. This heuristic can lead to a false sense of security, as attackers can easily replicate these visual indicators to create legitimate phishing sites. While this example illustrates how attackers can exploit representativeness bias in users to gain access to sensitive information, the research presented here investigates how representativeness bias in attackers can be used to create more effective cyber defenses.
One facet of representativeness bias, sample size insensitivity (SSI), accrues when people overlook the importance of the sample size on which their mental model is built. To this end, people will frequently give undue weight to conclusions based on small sample sizes when drawing conclusions about large, representative populations \cite{HamillEtAl80, TverskyKahneman82, GertnerEtAl16}. 
Results concerning SSI in our experiments are explained in Section~\ref{sec:methodology}.

We bring to the attention of the reader that while cyber data are commonly available during a cyberattack, direct indices of biases underlying an attacker's behavior are not. We organized a series of capture-the-flag (CTF) events to simulate and capture real-world adversarial behavior. In the CTFs, participants of different levels of expertise were invited to test their skills against a series of challenges carefully designed to manipulate cognitive biases, particularly representativeness. We ran these CTF exercises at two well-known international cyber-security conferences, namely the Hack In The Box\footnote{\url{https://conference.hitb.org/hitbsecconf2024bkk/saikoctf/}} (HITB) and the European Cybersecurity Challenge\footnote{\url{https://ecsc2024.it/saikoctf}} (ECSC) conferences, respectively.

Our experiments hypothesized that cyber defense strategies would affect an attacker's propensity to select certain (biased) action pathways. This approach assumes that psychology-informed defenses will negatively impact the attacker's behavior. Thus, the goal of this research was to show how an attacker's success could be impeded by carefully designed defenses that lead to a waste of resources, including increased expenditure of time and effort on the task at hand (i.e., capture-the-flag activities). 

Our results demonstrate a significant affinity for pathways triggering representativeness bias, with participants in treatment groups (i.e., exposed to the cyber defense tailored to representativeness bias, the so-called \textit{bias trigger}) spending $30.96$ seconds on non-vulnerable sites compared to $9.54$ seconds in control groups (i.e., not exposed to the bias trigger) ($p=0.016$). This work underscores the potential of psychology-informed defenses to enhance cyber resilience.

Our contributions can be summarized as follows:
\begin{itemize}
    \item We identify and emphasize the impact and potential of psychology-informed defense mechanisms in thwarting adversarial threats.
   \item We provide the results of a first-of-its-kind study that relies on CTF events to simulate and capture not only the real-world implications of an adversarial attack but also the cognitive biases of the participants while addressing the CTF challenges.
   \item We present our preliminary results discussing the impact and influence of representativeness bias in the attacker's decision-making processes.
   \item We discuss the impact and implications of our work for future research in the field, showing how this could radically change how cyber threats are addressed.
\end{itemize}

The rest of this paper is organized as follows: Section~\ref{sec:relatedWord} discusses the current state of the art in this field and its limitations. The methodology that we used is discussed in Section~\ref{sec:methodology}, followed by a brief description of the CTF instrumentation and data collection mechanisms employed, Section~\ref{sec:instrumentation}. Section~\ref{sec:analysis} analyzes our experiments manipulating representativeness bias, where we consider paths selected during the CTF challenges and the amount of time spent on specific challenges and tasks as key outcome variables. Details about the ethical considerations related to this work and respective IRB protocols are provided in Section~\ref{sec:ethical}. We provide concluding observations and suggestions for future work in Section~\ref{sec:conclusions}.

\section{Related Work}
\label{sec:relatedWord}

Cyber attackers exhibit a variety of standard decision-making biases, which have recently been captured in the social science literature \cite{gutzwiller2019cyber, JohnsonEtAl21}. Additional Cognitive Vulnerabilities (CogVulns) include the attacker’s culture, personality traits, emotional states, and cyberpsychology attributes. These factors shape a hacker’s behavior over time based on their state and dispositional factors. For example, a hacktivist trying to find damaging information about a company might become more loss-averse (a cognitive bias) after discovering the “smoking gun” that shows corporate indiscretion. This same hacktivist may have this bias dampened or elevated depending on where they stand on personality attributes such as Emotional Stability and Agreeableness. 

Cyber defenses based on deception (that is, honey-x strategies and decoys) have been developed for more than 20-years by researchers \cite{PanEtAl16, GuEtAl07, Spitzner03}, and these research concepts are coming to market~\cite{Proofpoint}, \cite{Alcalvio}. However, such products often require pre-defined configurations with static deployments that help attackers identify and avoid such defenses. 
Deception-based defenses can use System 2 thinking (i.e., slow effortful decision-making not linked to biases) and have been commercially available for years (e.g., Proofpoint, formerly Illusive), but these tools do not track and exploit the attacker's cognitive and/or emotional state.

Understanding adversarial cognitive state may provide richer attack and defense surfaces, particularly against Advanced Persistent Threats (APTs), whose campaigns are methodical, adaptive, and deeply embedded in strategic behavior. To investigate these complex psychological dynamics, both studies that test participants in longer experiments and competitive, time-limited simulations like CTF exercises are essential. Experiments, such as the Tularosa Study, offer controlled environments, in which the impact of deception on attacker cognition, behavior, and physiology can be rigorously measured (e.g., galvanic skin response, cognitive battery performance) using professional red teamers as proxies for real adversaries who are exposed to the tasks for extended time periods \cite{ferguson2018tularosa}. These studies validate that the presence of cyber deception, especially when paired with psychological manipulation (i.e., merely informing adversaries that deception may be present), can affect the attacker performance and forward progress (e.g., delayed exploitation, increased interaction with decoys), often exploiting cognitive biases like the sunk cost fallacy or confirmation bias~\cite{ferguson2021examining}.

CTF exercises can add a variation of ecological validity by introducing competitive pressure, variable adversary goals and, in certain scenarios, the emergent behaviors of team dynamics. In addition, each \saikoctf\footnote{\url{https://saikoctf.org/}} challenge is carefully designed to control dependent and independent variables to test bias-specific hypotheses. These settings foster more naturalistic decision-making under uncertainty and, when appropriately instrumented, can complement controlled studies by revealing real-time cognitive and emotional responses in unstructured, goal-driven environments~\cite{ferguson2018tularosa}. Ideally, both approaches can assist cyber defenders to better model, predict, and manipulate adversarial mental models.

Researchers have begun to focus on Moving Target Defenses (MTD) as a disruptive cyber defense solution \cite{ZhuangEtAl14, JajodiaElAl11}. MTDs create more complex, unpredictable attack surfaces, significantly bolstering system resilience. Some researchers have already begun to integrate MTD and honey-x techniques \cite{AchleitnerEtAl16, pagnotta2023dolos}. The research presented here uses a cyber psychology-informed approach to take these techniques to a new level. It exploits learned attacker cognitive biases to strategically manage deception campaigns that are designed for maximum attacker impact while using MTD concepts to dodge and skirt around attacks. This is done by intelligently adapting and orchestrating defenses that control network and host resources. 

This begs the question:~\emph{Why might such an approach be especially effective?} Cognitive biases are known to affect human decision-making and performance. Kahneman and Tversky's seminal work in this domain \cite{Kahneman11} led not only to their sharing a Nobel Prize in economics, but also to a whole new approach to understanding human decision-making under conditions of uncertainty. Much of the work in psychology and economics has focused on bias mitigation (reducing the influence of problems in decision-making); the current approach turns this idea on its head: Enhancing biases will lead to deleterious performance in cyber adversaries. That is, given the continuum underlying these models -- biases to heuristics -- we can leverage social science research to construct sophisticated defense strategies.
%

\section{Experiment Methodology}
\label{sec:methodology}

Our research investigates how CogVulns can be exploited to enhance cyber defenses. Specifically, we focus on representativeness bias, a common cognitive shortcut affecting probability estimates.
We conducted experiments using capture-the-flag (CTF) events to simulate real-world adversarial behavior. Participants were exposed to scenarios designed to trigger representativeness bias, allowing us to measure its impact on decision-making. The cyber defense designed to elicit a biased reaction from the participants is referred to as a \emph{bias trigger} throughout the remainder of this paper. 

\subsection{Experiment Setup}\label{ssec:exp-setup}
Our \saikoctf
experiments were held at several hacker conferences to attract individuals with cyber hacking interests and skills. Similar CTF events are common during hacking events so participants are well versed in what to expect from them. Different biases were investigated at various conferences. Representativeness bias was tested at two international conferences: The Hack In The Box (HITB) Security Conference in Bangkok, Thailand, and the European Cybersecurity Challenge (ECSC) in Turin, Italy. Participants signed up for an in-person CTF event at HITB and ECSC that lasted about two hours for each participant. 

Human Subject Research (HSR) is subject to federal laws, and interested individuals were informed of their rights and given information about the research objectives, approach, and methodology as part of the consenting process. Only consented participants were allowed to participate in \saikoctf. To protect participants' privacy, no personally identifiable information, such as name, email, or hacker handles, was collected. Interested hackers registered via a URL, where they were assigned randomly generated hacker handles and passwords. Those credentials authorized access to the cyber range that hosted the CTF challenges. The handles were also used to tag the data collected and publish results on the leaderboard. 

Participants were given six CTF challenges. They were timed remote service exploitation, web applications, and password cracking challenges. Before participating in these challenges, they completed a socio-demographic questionnaire (i.e., items asking of their gender, age, and country of origin). Interleaved within each challenge were other assessments measuring affective state, cognitive biases, cognitive aptitude, and personality. The fullness of analysis of all these assessments in relation to the cyber behavior collected during \saikoctf is subject to further exploration. This paper focuses on the analysis of the effectiveness of bias triggers.

The participants worked on a Kali Linux virtual machine (VM) without Internet access. Pen-testing tools, documentation, and man pages were provided as part of the Kali Linux VM so all participants could access the same tools and information. Participants hunted for and captured specified flags as fast as possible in the time available to win challenge prizes. 
The Kali Linux VM was instrumented to collect data to analyze the effectiveness of our bias triggers. See Section~\ref{sec:instrumentation} for details on the collected data and Section~\ref{sec:analysis} for the experimental analysis approach and results. 

Among the six challenges that each participant performed, two challenges contained cyber defenses manipulating the sample size insensitivity (SSI) facet of representativeness bias. For the remainder of this paper, we refer to those CTF challenges as \repa and \repb. The details of the bias trigger (that is, cyber defense technique used to elicit a measurable effect on the participants' cyber attack behavior) that were tested in \repa and \repb are given in Section~\ref{sec:challengeDesign}. 

The experiment used a within-subjects design, meaning each participant was exposed to the experiment's control and treatment conditions. In the control condition, participants were not exposed to the bias trigger (the experiment's independent variable). However, in the treatment condition, participants were exposed to the bias trigger. In a within-subject experiment that uses CTF challenges, a hacker has to play a CTF challenge in both ways, with and without a bias trigger. 

This constitutes a problem because if a hacker sees the same challenge twice and has solved it on the first pass, it will influence their approach the second time due to the learning effect. For this reason, we designed two challenges, \repa and \repb, that were different in appearance but required the same techniques for solving them. The bias trigger in the treatment versions of \repa and \repb uses the same underlying defense mechanism. Thus, we designed and implemented a total of four CTF challenges, namely \repa-Control, \repa-Treatment, \repb-Control, and \repb-Treatment. 

Furthermore, the design is counter-balanced to control for potential {\it order effects} where the sequence of the treatment could influence the results, not the treatment itself. In practice, this means that participants are separated into two groups, A and B, and the order in which they are exposed to control and treatment versions of the challenges is alternating (see Table~\ref{tab:within-subjectDesign}). Table~\ref{tab:no-participants} provides the number of participants in each event corresponding to each challenge and group (control / treatment).

\begin{table}[]
    \caption{Counter-balanced, within-subjects experiment design using two pairs of CTF challenges, where a pair constitutes a control and treatment version of a challenge.}
    \centering
    \begin{tabular}{c|c}
        \hline
       Group A  & Group B \\
       \hline
        \repa-Control & \repa-Treatment\\
        \repb-Treatment & \repb-Control \\
    \end{tabular}
    \label{tab:within-subjectDesign}
\end{table}

\begin{table}[h]
\centering
\caption{Number of participants present in each event, separated by challenge as well as control and treatment groups, respectively. Note that the participants could opt out at any time, thus the discrepancies between control and treatment participant numbers.}
\label{tab:no-participants}
\begin{tabular}{|c|cccc|}
\hline
\multirow{3}{*}{\textbf{Event}} & \multicolumn{4}{c|}{\textbf{Challenges}} \\ \cline{2-5} 
 & \multicolumn{2}{c|}{\textbf{Rep1a}} & \multicolumn{2}{c|}{\textbf{Rep1b}} \\ \cline{2-5} 
 & \multicolumn{1}{c|}{\textbf{Control}} & \multicolumn{1}{c|}{\textbf{Treatment}} & \multicolumn{1}{c|}{\textbf{Control}} & \textbf{Treatment} \\ \hline
\textbf{ECSC} & \multicolumn{1}{c|}{28} & \multicolumn{1}{c|}{30} & \multicolumn{1}{c|}{26} & 29 \\ \hline
\textbf{HITB} & \multicolumn{1}{c|}{9} & \multicolumn{1}{c|}{7} & \multicolumn{1}{c|}{9} & 7 \\ \hline
\end{tabular}
\end{table}
\subsection{Challenge Design}\label{sec:challengeDesign}

The bias triggers for \repa and \repb were designed as follows. Both challenges were website exploitation challenges with the same hypothesis grounded in the SSI of representativeness bias: a hacker who sees mentions of a service without any other supporting evidence is more likely to target that service. 

Participating hackers were instructed that they had infiltrated the \saikoctf, LLC enterprise network and needed to find a flag in the webserver at \texttt{intranet.}\saikoctf\texttt{.org} in the root, but the flag was not directly accessible from the root. 

\begin{figure}[ht]
    \centering
    \begin{minted}[
          autogobble,
          frame=single,
          framesep=1mm,
          baselinestretch=1.2,
          breaklines,
          fontsize=\scriptsize,
        ]{bash}      
Log Aggregation Server: LogHub 
Version: 2.1.0 
Hostname: loghub-server-01 
Environment: Production 
Aggregation Method: Real-time streaming 
Data Source: System and Web Server Logs 
Aggregation Period: Last 24 hours 
Purpose: Monitoring and analysis of system performance and security incidents 

[error] [127.0.0.1] unknown: /upload.php 
[error] [127.0.0.1] unknown: /admin.php 
[info] [127.0.0.1] status installed mysql-client-5.0 5.0.32-7etch5 
[info] [127.0.0.1] status half-configured zlib1g-dev 1:1.2.3-13 
[error] [127.0.0.1] unknown: /admin.php 
[error] [127.0.0.1] unknown: /admin.php 
[error] [127.0.0.1] unknown: /upload.php 
[info] [127.0.0.1] status installed sysklogd 1.4.1-18 
[info] [127.0.0.1] status config-files nmap 4.11-1 
[error] [127.0.0.1] unknown: /admin.php 
[error] [127.0.0.1] unknown: /admin.php 
[error] [127.0.0.1] unknown: /admin.php 
[error] [127.0.0.1] unknown: /admin.php 
[info] [127.0.0.1] status not-installed sysklogd <none> 
[error] [127.0.0.1] unknown: /admin.php 
[error] [127.0.0.1] unknown: /admin.php 
[error] [127.0.0.1] unknown: /admin.php 
[error] [127.0.0.1] unknown: /admin.php 
[error] [127.0.0.1]
    \end{minted}
    \caption{Content of logs shown in the treatment version of \repa. In this challenge, the control group did not have any logs.}
    \label{fig:treatment-snippet}
\end{figure}

\textbf{\repa Design:}
In the \repa CTF scenario, the bias trigger is implemented as follows: 
\begin{itemize}
    \item Participants in the treatment group are presented with numerous log file entries that reference the non-vulnerable endpoint (the \texttt{/admin} endpoint).
    \item Participants in the control group do not have any log files.
\end{itemize}

The treatment version had log entries from a simulated log aggregator that showed errors and miscellaneous information messages. Error messages only showed ``\texttt{unknown}'' with no specifics. Miscellaneous information messages included messages designed to throw people off course. For example, we hypothesized that mentioning backups could cause participants to try to probe for a backup service. 
Log content was static, such that all participants received the same content. The log message distribution was: 
\begin{itemize}
    \item 50\% alerts for the non-vulnerable \texttt{/admin.php} endpoint 
    \item 10\% alerts for the vulnerable \texttt{/upload.php} endpoint
    \item 40\% miscellaneous info messages.  
\end{itemize}
A snippet of the content shown during the treatment version is depicted in Figure~\ref{fig:treatment-snippet}.

\begin{figure}[t]
\centering
\includegraphics[width=\columnwidth]{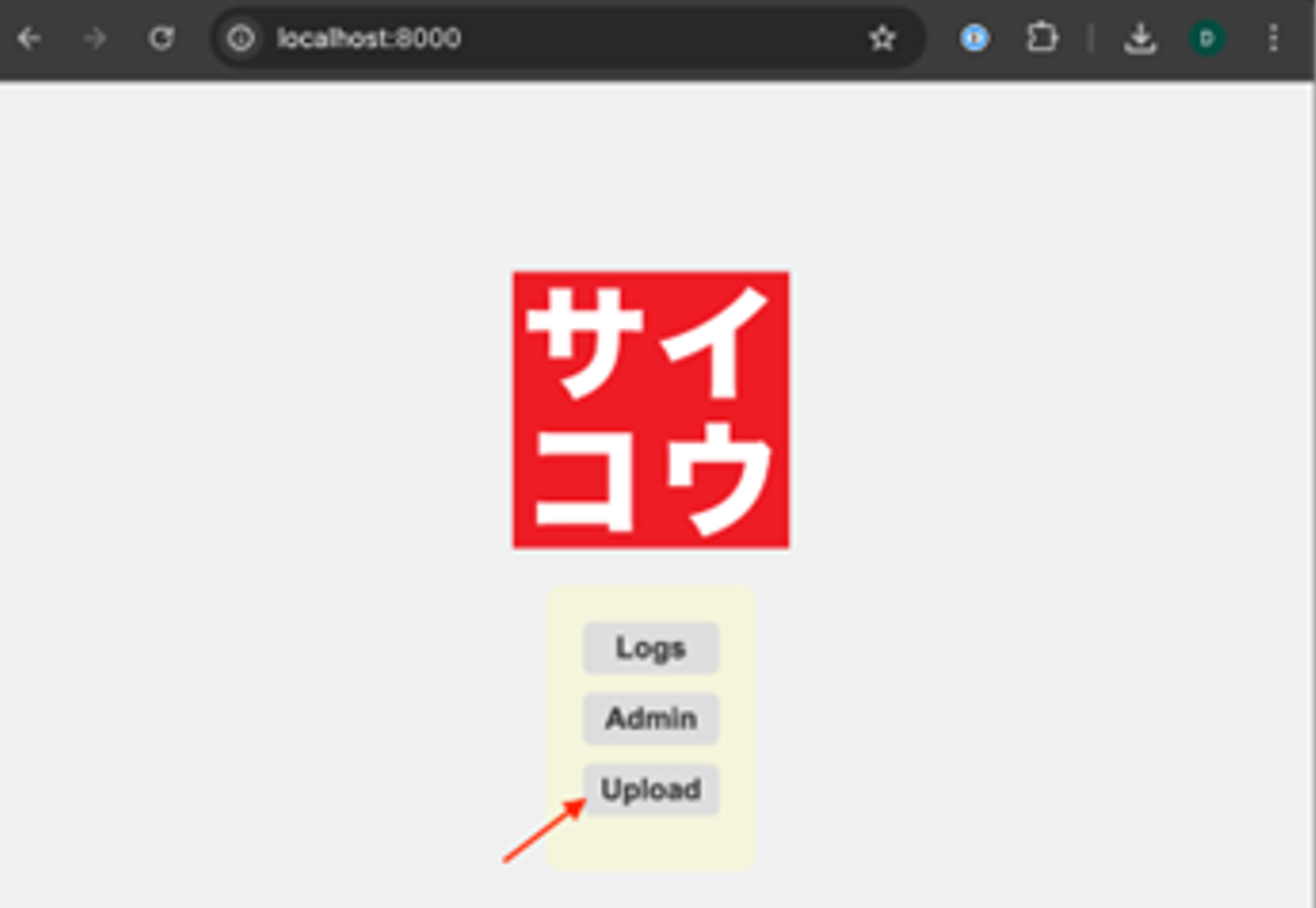}
\caption{\saikoctf index page for \repa challenge~}
\label{fig:saiko-rep}
\end{figure} 

\begin{figure*}[t]
\begin{subfigure}[t!]{0.55\textwidth}
    \centering
    \begin{minted}[
          autogobble,
          frame=single,
          framesep=1mm,
          baselinestretch=1.2,
          breaklines,
          fontsize=\scriptsize,
        ]{bash}      
Log Aggregation Server: LogHub 
Version: 2.1.0 
Hostname: loghub-server-01 
Environment: Production 
Aggregation Method: Real-time streaming 
Data Source: System and Web Server Logs 
Aggregation Period: Last 24 hours 
Purpose: Monitoring and analysis of system performance and security incidents 

[info] [127.0.0.1] status half-configured mysql-common 5.0.32-7etch5 
[info] [127.0.0.1] Lab14 kernel: usb 3-1.2: not running at top speed; connect to a high speed hub 
[info] [127.0.0.1] status unpacked libplrpc-perl 0.2017-1.1 
[info] [127.0.0.1] status half-configured sysklogd 1.4.1-18 
[info] [127.0.0.1] status not-installed nmap <none> 
[info] [127.0.0.1] status unpacked libplrpc-perl 0.2017-1.1 
[info] [127.0.0.1]
    \end{minted}
\end{subfigure}
\begin{subfigure}[t!]{0.45\textwidth}
\centering
    \begin{minted}[
          autogobble,
          frame=single,
          framesep=1mm,
          baselinestretch=1.2,
          breaklines,
          fontsize=\scriptsize,
        ]{bash}
Log Aggregation Server: LogHub 
Version: 2.1.0 
Hostname: loghub-server-01 
Environment: Production 
Aggregation Method: Real-time streaming 
Data Source: System and Web Server Logs 
Aggregation Period: Last 24 hours 
Purpose: Monitoring and analysis of system performance and security incidents 

[error] [127.0.0.1] unknown: /cat.php 
[error] [127.0.0.1] unknown: /cat.php 
[error] [127.0.0.1] unknown: /cat.php 
[info] [127.0.0.1] status installed libdbd-mysql-perl 3.0008-1 
[error] [127.0.0.1] unknown: /cat.php 
[error] [127.0.0.1] unknown: /dog.php 
[error] [127.0.0.1] unknown: /cat.php 
[error] [127.0.0.1] 
  \end{minted}
\end{subfigure}%
\caption{Content of logs shown in CTF challenge \repb in the control (left) and the treatment version (right).}
\label{fig:repb-snippet}
\end{figure*}

Figure~\ref{fig:saiko-rep} shows \repa's top-level web page. We hypothesized that hackers would engage with the admin sub-page in the treatment version because 50\% of the log entries mention \texttt{admin.php}. However, the solution was to navigate to the file upload endpoint from the index page and upload a PHP web shell with a file extension other than lowercase ``.php'' to bypass the simple input filter. Hackers could then visit the web shell URL and print the flag. 

\repa was designed to have a very simple test of representativeness by examining how a participant's behavior will be impacted from being given prior information on endpoints in the environment relative to no prior knowledge of the environment. With the treatment group having logs of the endpoints and over-representation of the non-vulnerable \texttt{/admin.php} relative to vulnerable \texttt{/upload.php}, we anticipated a clean comparison compared to the control group having no prior knowledge on endpoint logs. Thus, the user will have an equal likelihood of initial examination of either endpoint absent positional effects, discussed further in Section \ref{sec:considerations}.  The choice of the endpoints \texttt{/admin.php} and \texttt{/upload.php} was made to be reflective of common endpoint labels seen in real deployed systems.

\textbf{\repb Design:}
\repb is similar to \repa in that two endpoints were provided to the participants. In \repb, the logs also showed errors and miscellaneous information messages from a simulated log aggregator. Error messages only showed “unknown” with no specifics. Miscellaneous information messages included some messages designed to throw people off course. 

In contrast to \repa, \repb the endpoints were labeled \texttt{/dog} and \texttt{/cat}. In the \repb CTF scenario, the trigger is implemented as follows: 
\begin{itemize}
    \item Participants in the treatment group are presented with a high amount of logs on the non-vulnerable endpoint (which is the \texttt{/cat} endpoint) and fewer mentions of the vulnerable endpoint (which corresponds to the \texttt{/dog} endpoint.)
    \item Participants in the control group have logs filled 100\% with irrelevant miscellaneous information. We hypothesized that attackers would initially target the two endpoints with equal probability. 
\end{itemize}

The treatment version of \repb showed logs that had the following distribution of 
\begin{itemize}
    \item 60\% alerts for the non-vulnerable \texttt{/cat} endpoint 
    \item 30\% alerts for the vulnerable \texttt{/dog} endpoint
    \item 10\% miscellaneous info messages.  
\end{itemize}

Our design choice in \repb sought to examine representativeness by examining how a participant behavior will be impacted from being given over-represented data of non-vulnerable to vulnerable (treatment) relative to no informative knowledge of logs (control). For the treatment condition, we hypothesized that hackers would initially target the non-vulnerable endpoint with greater frequency. 
Snippets of the content shown during the control and treatment version of \repb are depicted in Figure~\ref{fig:repb-snippet}.

\section{Instrumentation for Data Collection}
\label{sec:instrumentation}

We instrumented a Kali Linux VM to be used by CTF participants for their attack activities. Instrumentation includes a keylogger, a terminal logger, and a web browsing logger. The keylogger captures and time-stamps all key presses, regardless of which window the participant is typing in. The terminal logger captures all characters typed and received in the terminal shell, along with timing data. This enables us to capture not only the commands entered but also the information displayed in response. 
In addition to the URLs requested, the weblogger captures and time-stamps all browser interactions using the keyboard and mouse. Instrumentation also captured all processes started, all content copied to the clipboard, and all commands started via the menu. Finally, we captured video screen recordings of the Kali VM for human-centric quality analysis and to provide ground truth for measuring automated data collection.

We instrumented \texttt{nginx/access.log} and \texttt{php.log} on the target to capture web activities. All web traffic was forced to use HTTP, and packet capture was used to collect the unencrypted network packets. Moreover, we incorporated additional instrumentation explicitly designed for login attempts. This data was used to measure other CTF challenges. Data is purposefully captured in redundant ways to provide backup and multiple perspectives to support bias signal analysis.

A post-processing data analysis pipeline integrates the different artifacts from the attacker’s activity to create a holistic picture of the attacker’s behavior.
Together, this information helps to illuminate the hacker’s cognitive decision-making processes and how they respond to different bias triggers. 
We analyzed the collected cyber data to provide metrics about trigger effects and user errors, extract typical workflows, and determine flag capture results.

\section{Analysis Approach and Results}
\label{sec:analysis}

\subsection{Path Selected}
Cyber data was analyzed for \repa and \repb to determine the significance and magnitude of the effect of the cognitive bias trigger. In both challenges, we developed analyses of the participant activity and choice during CTF execution with analyses built on the chi-square testing model.

To this end, our chi-square analysis tested for a difference in whether or not the participant selected a choice coded as `biased' at the position of a cognitive vulnerability trigger, e.g., the choice of endpoint to interact with/exploit during CTF. The determination of a `biased' action was whether or not the participant chose the `biased' coded endpoint—graded identically on control and treatment to establish if there was a significant difference from the base rate choice. 

\begin{figure}[htb]
\centering
\begin{subfigure}{.5\columnwidth}
    \centering
    \includegraphics[width=\textwidth]{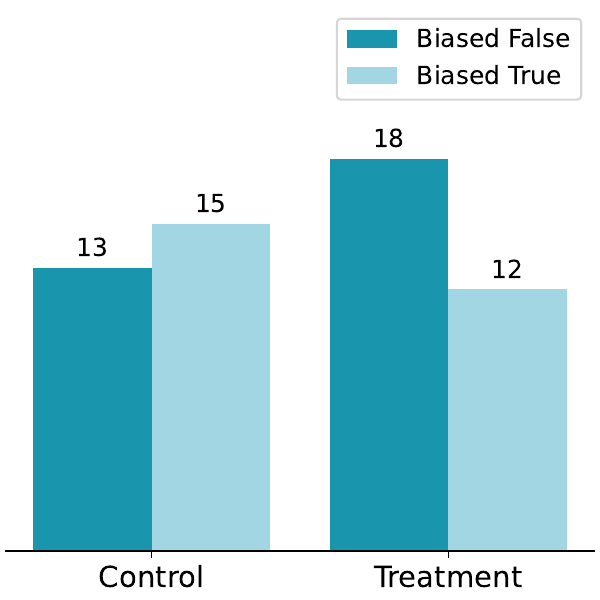}
    \caption{$E = 0.139$;~$P = 0.440$}
\end{subfigure}%
\begin{subfigure}{.5\columnwidth}
    \centering
    \includegraphics[width=\textwidth]{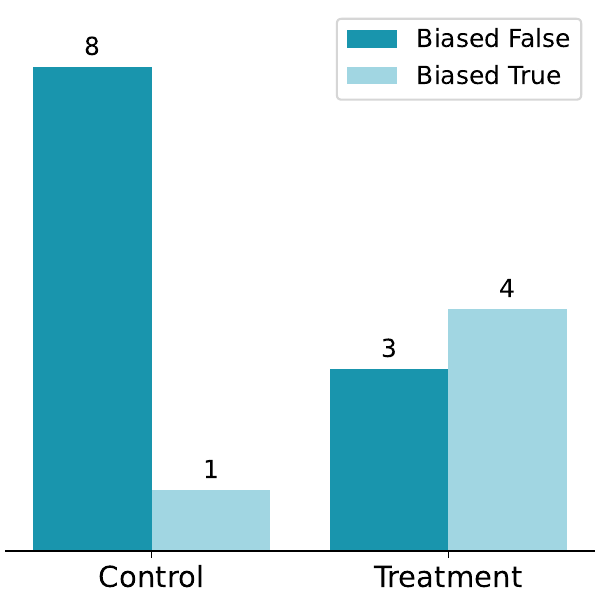}
    \caption{$E = 0.638$;~$P = 0.154$}
\end{subfigure}
\caption[short]{\repa results for ECSC (left) and HITB (right) events. Effect size ($E$) and respective P-values ($P$) are shown for each plot.}
\label{fig:chi-rep8a}
\end{figure}

\begin{figure}[htb]
\centering
\begin{subfigure}{.5\columnwidth}
    \centering
    \includegraphics[width=\textwidth]{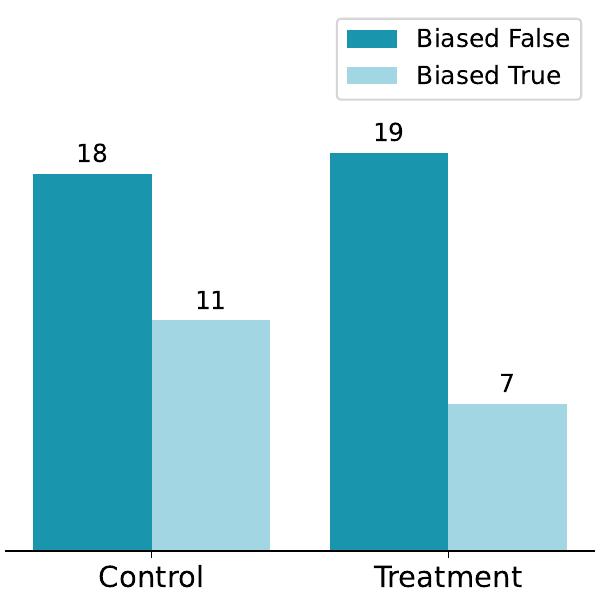}
    \caption{$E = 0.095$;~$P = 0.561$}
\end{subfigure}%
\begin{subfigure}{.5\columnwidth}
    \centering
    \includegraphics[width=\textwidth]{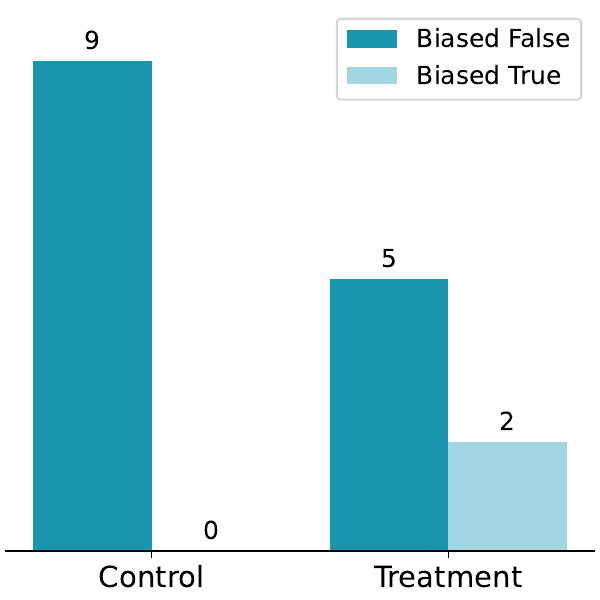}
    \caption{$E = 0.673$;~$P = 0.341$}
\end{subfigure}
\caption[short]{\repb results for ECSC (left) and HITB (right) events. Effect size ($E$) and respective P-values ($P$) are shown for each plot.}
\label{fig:chi-rep8b}
\end{figure}

Our results for both \repa and \repb scenarios across ECSC and HITB conferences are shown in Figures~\ref{fig:chi-rep8a}-\ref{fig:chi-rep8b}. For \repa, we saw no significance in participants' first choices when comparing control and treatment selections across individuals at either ECSC or HITB. Similarly, in \repb, we saw no significance in participants' first choices when comparing control and treatment selections across individuals at either ECSC or HITB. 

This first pass at analysis was unsuccessful in determining strong effects from the condition (Treatment vs. Control) within the challenges as a binary measure. Instead, our subsequent modeling efforts focused on continuous measures of the activity exhibited by the participants in the challenges.  

\subsection{Average Time Spent on Challenges}
Beyond initial path choice analyses, we took an additional perspective on the analysis of the representativeness trigger bias in \repa and \repb scenarios based on time spent in a challenge. More specifically, we examined whether there were any notable differences in the average time spent on vulnerable vs non-vulnerable paths for both control and treatment groups. As part of our analysis, we found two settings with potentially statistically significant differences between the control and treatment groups, both cases related to the \repb event in both HITB and ECSC events. 

\begin{figure}[htb]
\centering
\begin{subfigure}{.5\columnwidth}
    \centering
    \includegraphics[width=\textwidth]{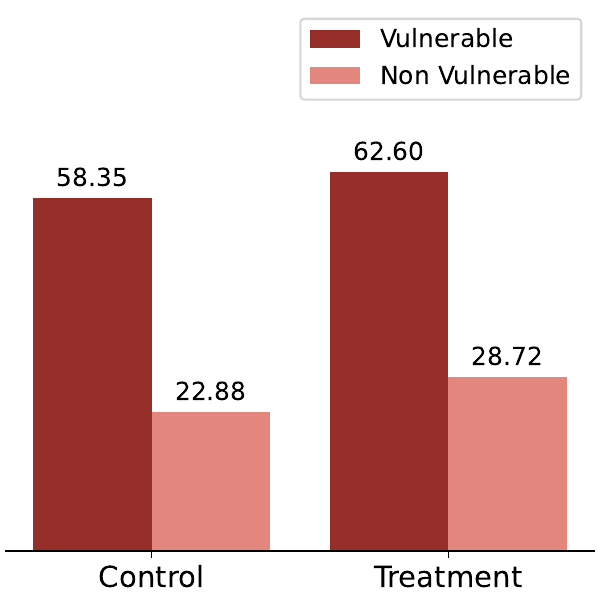}
    \caption{ECSC}
\end{subfigure}%
\begin{subfigure}{.5\columnwidth}
    \centering
    \includegraphics[width=\textwidth]{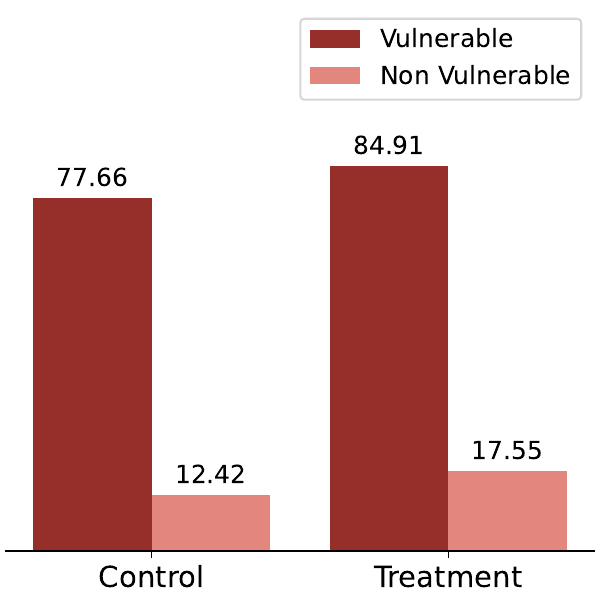}
    \caption{HITB}
\end{subfigure}
\caption[short]{\repa results for ECSC (left) and HITB (right) events.}
\label{fig:event-rep8a}
\end{figure}

\begin{figure}[htb]
\centering
\begin{subfigure}{.5\columnwidth}
    \centering
    \includegraphics[width=\textwidth]{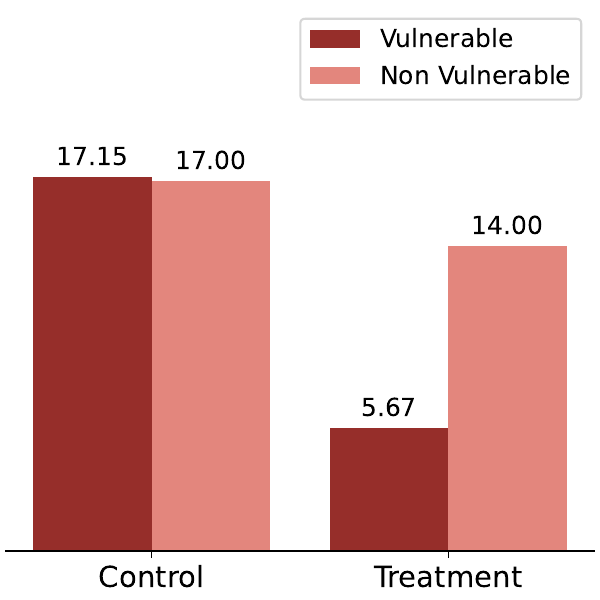}
    \caption{ECSC}
\end{subfigure}%
\begin{subfigure}{.5\columnwidth}
    \centering
    \includegraphics[width=\textwidth]{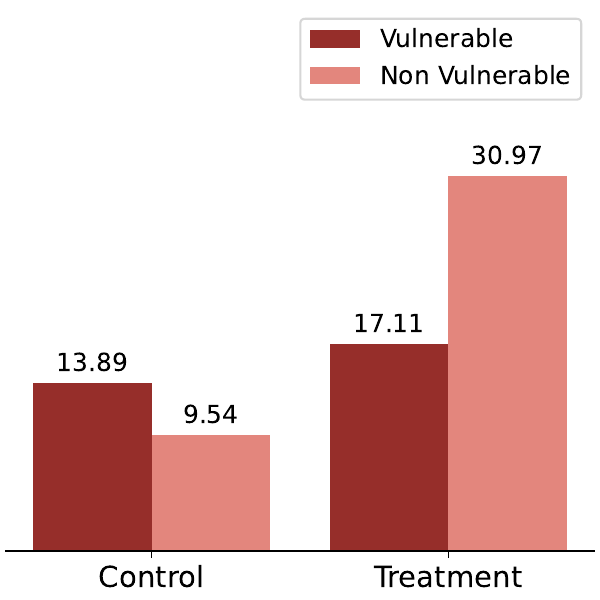}
    \caption{HITB}
\end{subfigure}
\caption[short]{\repb results for ECSC (left) and HITB (right) events.}
\label{fig:event-rep8b}
\end{figure}

To answer the above question, we focused on collected web log data, namely records of browser activity logs for each participant. This data was further complemented with metadata information collected during the CTF events, such as whether a participant was part of the control or treatment group in each challenge. The weblogger data was sorted by time. Then, the data was filtered by event, challenge, and treatment group. For each participant in a given group (treatment/control), we calculated the time (in seconds) spent on a particular site, where the initial time (i.e., \texttt{start time}) corresponds to the first instance the participant is on a given site, and the end time corresponds to the moment in time when the site URL changes, i.e., the participant switches to another site. This procedure resulted in a data frame containing records of the time spent (i.e., \texttt{duration}) on a site, recording the duration as separate records for different visits to the same site. We then calculated the average time spent on vulnerable and non-vulnerable sites, the results of which are depicted in Figures~\ref{fig:event-rep8a}-\ref{fig:event-rep8b}.
It can be noted immediately from the plots in Figures~\ref{fig:event-rep8a}-\ref{fig:event-rep8b} that there are some differences between control and treatment group actions in \repb challenges taking place in both HITB and ECSC events. We then defined different hypotheses and calculated $t$ and $p$-values in each case.

During our analysis of HITB results, we found a potentially significant difference between control and treatment groups on the average time spent on non-vulnerable sites for the \repb challenge, Figure~\ref{fig:event-rep8b} reported as mean $\pm$ standard deviation. More specifically, treatment group spent ($30.97 \pm 45.71$) seconds compared to the control group ($9.54 \pm 14.95$ seconds), with $t= -2.49$, $p= 0.016$. Whereas our analysis of the ECSC cyber data showed that the treatment group spent significantly less time on vulnerable sites ($5.67 \pm 8.33$ seconds) compared to the control group ($17.15 \pm 26.54$ seconds), with $t = 2.69$, and $p=0.008$ this again when evaluating the \repb data, Figure~\ref{fig:event-rep8b}. While these results are promising, given the high fluctuations in the standard deviation values (indicating non-normal distributions) found during the web logger data analysis, as part of the next steps, we will seek to verify and complement the analysis with cues and signals originating from other forms of data collected during the respective events. 

We also modeled the average time spent within a linear mixed-effects model to control the variance in point estimation in the above analysis. Treating the event (HITB, ECSC), challenge (\repa, \repb), and condition (treatment, control) as fixed effects and the by-group (participant) slopes as random effects. We chose this method to directly control for the hierarchical structure of the data (e.g., participants within conferences) and control for correlated observations within participants (e.g., a given participant will see \repa and \repb sequentially within a conference but be switched within the condition between challenges). 

From our modeling, shown in Table~\ref{tab:avgtime-table-lme}, we found significance within intercepts, event, challenge, and condition $p< 0.000$ with absolute values of the z-scores for fixed effects being $15.987$, $32.119$, and $4.738$ in event, challenge, and condition, respectively. These results indicate significant differences in time on vulnerable paths as a function of the event, challenge, and condition. However, as evidenced by the earlier analyses of the means, the effects are not as expected from our initial experimental design.

\begin{table}[]
    \caption{Results of linear mixed effects modeling of time spent on a vulnerable path for treatment and control groups in HITB and ECSC events for Rep1a and Rep1b.}
    \centering
    \begin{tabular}{c|c|c}
        \hline
       Effect  & $|z|$ & $p$ \\
       \hline
        Intercept & $56.490$ & $<0.000$ \\
        Event & $15.987$ & $<0.000$ \\
        Challenge & $32.119$ & $<0.000$ \\
        Condition & $4.738$ & $<0.000$ 
    \end{tabular}
    \label{tab:avgtime-table-lme}
\end{table}

Considering the results discussed herein, we have an interesting, if slightly contradictory, story from initial analyses. For example, there are significant results for the time spent on the vulnerable path by condition, challenge, and conference, see Figure~\ref{fig:event-rep8b} and Table~\ref{tab:avgtime-table-lme}. However, we find perplexing results when considering the representativeness bias triggering in \repa and \repb - the triggering of performers in \repb caused them to pursue the non-vulnerable path significantly more in the treatment group, which was the opposite of expectations from the design.
Additionally, when examining the linear mixed effects model in Table~\ref{tab:avgtime-table-lme}, we see unanticipated results. We expected to see strong effect sizes related to the condition, but not the challenge, yet both are evident. This suggests that the populations being sampled differed more drastically than anticipated from our initial experimental design and recruitment plans. 
After reviewing the scenario designs, we believe that the \repa challenge had significant differences in the control and treatment for the lack of depth in information compared to the \repb challenge with the depth of information in logs being held similarly between conditions. This would support our observation of challenge being a strong observed effect in the linear mixed effect modeling. 
Further modeling efforts will include examining for positional bias and incorporating richer data on the performers as collected in surveys and physiological measures. We anticipate with additional data and modeling a more complete understanding of the participant behavior during CTF exercises will emerge.

\section{Ethical Considerations}
\label{sec:ethical}

All researchers who conduct studies using human participants are bound by
professional ethical standards for the conduct of such research. These standards
are mirrored in the rights that are guaranteed to research participants by federal
law (NIH regulation 45-CFR-46) (see also the Menlo Report \cite{kenneally2012menlo}). 
Before conducting the research study at conferences, our team submitted all the necessary documentation related to the HSR experiments taking place in both the conferences mentioned in this paper, namely ECSC and HITB. The Institutional Review Board (IRB) of Florida Institute for Human Machine Cognition (IHMC)\footnote{While not involved with the data evaluation and analysis procedures resulting in this work, we bring to the attention of the reader that IHMC is a member of the larger team behind this effort.}, through an IRB Authorization Agreement with SRI International, reviewed and approved the procedures corresponding to the protocol number:~\textbf{IRB-2024-0076}, in particular, \textbf{IRB-2024-0076:001988 (ECSC)} and \textbf{IRB-2024-0076:001985 (HITB)}.

As noted also in Section~\ref{ssec:exp-setup}, our team took all the necessary steps to provide the participants with details about the experiments. Participants were provided with an informed consent form. The consent gave an overview of the project, explaining the purpose of the research and the general approach and methodology, including a detailed schedule of the activities during participation. The consent form also provided detailed information on potential risks and risk mitigations. Participants were provided with information about how the collected data was kept confidential. Furthermore, the participants could decline consent or withdraw from the study at any time. They could also exercise their rights after the study, such as requesting that all the data they provided be deleted. These procedures ensured that they never lost their ability to make an autonomous decision.

Participants received a \saikoctf electronic badge (that included some digital puzzles) for participating. Moreover, it is common practice for hacker conferences to provide electronic badges or prize money. Individuals who achieved the top three \saikoctf scores received digital gift cards ($1^{st}$ place = USD 400; $2^{nd}$ place = USD 300; and $3^{rd}$ place = USD 100). Participants who completed the personality
assessment before the end of the conference also received a gift
card for USD 25. 

We believe that the research presented here will help spearhead novel research on psychology-informed active defense strategies, giving the defenders an edge against ever-evolving adversary threats.

\section{Conclusions}
\label{sec:conclusions}

\subsection{Considerations on Current Study}
\label{sec:considerations}
This study is an initial step toward investigating psychology-informed, dynamic cyber defenses. We studied the effectiveness of bias triggers exploiting the representativeness bias. Other bias triggers are currently under investigation. However, bias triggers are only one side of the coin towards novel, adaptive cyber defenses. The other side corresponds to the development of the so-called {\it bias sensors}. 

Bias sensors dynamically correlate robust patterns of cyber data as surrogates for CogVulns. 
For instance, would it be possible to identify which attack strategies explain an attacker's propensity to other biases such as loss aversion?
Let's assume an attacker has gained partial access through password-cracking strategies and has stopped progressing but continues down the same attack vector while risking their activities being more likely to be detected. Such attacker behavior might indicate their tendency for loss aversion. Once a bias sensor attributes a CogVuln to an attacker, a corresponding bias trigger targeting that CogVuln can be dynamically deployed.

After analysis of the data we found additional items that would have been optimal to control in a larger study (e.g., positional bias in \repa). Additionally, due to the short nature of these CTF exercises, the longer persistent behavior studied in experiments like the Tularosa study~\cite{ferguson2018tularosa} was not observed. This brings into question the direct applicability of some of these results for longer APT-like campaigns due to the forced time nature in CTFs including the ones developed herein. 

Additionally, the counterintuitive results in Section \ref{sec:analysis} show that more study is needed for clarity of the effect of our bias triggers in CTFs. The unexpected results of \repb treatment group pursuing non-vulnerable paths in the face of over-represented prior knowledge in the form of log data suggests latent factors may be at play that are not being controlled in the experimental design or currently modeled.

Regardless of the limitations for the current study, we recognize our work in validating the capability of bias triggers in potentially directing adversarial behavior even within a confined CTF as seen in Figure \ref{fig:event-rep8b} and Table \ref{tab:avgtime-table-lme}. 

\subsection{Future Work}
\label{sec:future-work}
Previous research has shown a significant survey of biases relevant to decision-making in cyber operations~\cite{johnson2020cyber}. In subsequent research, we will continue developing manipulations of several of these (i.e., confirmation bias, anchoring bias, loss aversion, overconfidence, and country-of-origin biases). Within the current work, we did not find significant differences in the initial path chosen upon exposure to the cognitive bias trigger, but we did identify potential significance in the time spent within the paths as a function of the trigger. This effect is practically meaningful: It can be used, for example, to purposely thwart an actor by elongating their time within a honeypot. 

This work will continue exploring additional biases in CTF environments like the one described above, as well as longer instantiations (e.g., online games) that better represent the conditions typically experienced by an attacker, improving so-called ecological validity. The bias manipulations per se require gold standards by which these can be compared and contrasted, so these studies will also include classic operationalizations of the various biases (e.g., assessment paradigms from the social sciences). Given a sufficient sample size, these assessments can lead to robust mediator and moderator analyses. These additional studies will also include a range of individual differences measures, such as personality, indicators of psychopathology, cognitive ability, and affective states. To understand hacker behavior more fully, we are also currently conducting studies where individual hacker differences are compared to those obtained from the general population.

It is our belief that validated bias triggers and sensors open the door to novel cyber defenses by augmenting traditional adaptive cyber defenses with psychology-informed defense mechanisms that take advantage of a new dimension: human attackers' behavioral vulnerabilities. 
%
\section*{Acknowledgments}

We would like to thank the anonymous reviewers and our shepherd for their valuable comments and feedback, which significantly improved our paper.
This research is based upon work supported in part by the Office of the Director of National Intelligence (ODNI), Intelligence Advanced Research Projects Activity (IARPA), via N66001-24-C-4505. The views and conclusions contained herein are those of the authors and should not be interpreted as necessarily representing the official policies, either expressed or implied, of ODNI, IARPA, or the U.S. Government. The U.S. Government is authorized to reproduce and distribute reprints for governmental purposes notwithstanding any copyright annotation therein.

\bibliographystyle{IEEEtran}
\bibliography{main}

\end{document}